\documentclass[aps,prd,twocolumn,superscriptaddress,showkeys]{revtex4}
\usepackage{amssymb,amsmath,graphicx,enumerate}
\usepackage{subfigure}
\usepackage{dcolumn,booktabs}
\usepackage[dvipdfm,colorlinks=true,citecolor=blue,pdfstartview=FitH]{hyperref}

\def\bea{\begin{equation}}
\def\eea{\end{equation}}
\newcommand{\rt}{Regge trajectory}
\newcommand{\rts}{Regge trajectories}
\newcommand{\bfr}{{\bf r}}

\newcommand{\bfpa}{{|\bf p|}}
\newcommand{\gev}{{\rm GeV}}

\newcommand{\sse}{spinless Salpeter equation}

\newcommand{\cltb}{$\bar{3}_c$}
\newcommand{\cltba}{\bar{3}_c}
\newcommand{\cls}{$6_c$}
\newcommand{\dqs}{$(qq')$}

\begin{document}
\title{Regge trajectories for the light diquarks}
\author{Jiao-Kai Chen}
\email{chenjk@sxnu.edu.cn, chenjkphy@outlook.com}
\affiliation{School of Physics and Information Engineering, Shanxi Normal University, Taiyuan 030031, China}
\author{Jia-Qi Xie}
\email{1462718751@qq.com}
\affiliation{School of Physics and Information Engineering, Shanxi Normal University, Taiyuan 030031, China}
\author{Xia Feng}
\email{sxsdwxfx@163.com}
\affiliation{School of Physics and Information Engineering, Shanxi Normal University, Taiyuan 030031, China}
\author{He Song}
\email{songhe\_22@163.com}
\affiliation{School of Physics and Information Engineering, Shanxi Normal University, Taiyuan 030031, China}
\date{\today}

\begin{abstract}
We attempt to present an unified description of the light meson spectra and the light diquark spectra by applying the Regge trajectory approach. However, we find that the direct application of the linear Regge trajectory formula for the light mesons and baryons fails. To address this issue, we fit the experimental data of light meson spectra and the light diquark spectra obtained by other theoretical approaches. By considering the light quark mass and the parameter $C$ in the Cornell potential, we provide a provisional Regge trajectory formula.
We also crudely estimate the masses of the light diquarks $(ud)$, $(us)$, and $(ss)$, and find that they agree with other theoretical results. The diquark Regge trajectory not only becomes a new and very simple approach for estimating the spectra of the light diquarks, but also can explicitly show the behavior of the masses with respect to $l$ or $n_r$.
Moreover, it is expected that the diquark Regge trajectory can provide a simple method for investigating the $\rho$-mode excitations of baryons, tetraquarks and pentaquarks containing diquarks.
\end{abstract}

\keywords{Regge trajectory, diquark, spectra}
\maketitle

%%%%%%%%%%%%%%%%%%%%%%%%%%%%%%%%%%%%%%%%%%%%%%%%%%%%%%%%%%%%%%%%%%%%%%%%%%

\section{Introduction}
Diquark correlations might play a material role in the formation of exotic tetraquarks and pentaquarks \cite{Jaffe:2004ph,Liu:2019zoy,Esposito:2016noz,Olsen:2017bmm,Guo:2019twa}.
Diquark substructure affects the static properties of baryons, tetraquarks and pentaquarks \cite{Gell-Mann:1964ewy,Lichtenberg:1967zz,
Anselmino:1992vg,Jaffe:2004ph,Barabanov:2020jvn,Selem:2006nd,Sonnenschein:2018fph,
Wilczek:2004im,Jaffe:2003sg,Brambilla:2019esw,Ida:1966ev,Lu:2017meb,Jaffe:1976ig,
Lebed:2015tna}.
%%%
The spectra of the light diquarks composed of two light quarks has been studied using various approaches. In Ref. \cite{Ebert:2007rn,Faustov:2021hjs,Faustov:2015eba}, the light diquark masses are calculated by using the Sch\"{o}dinger-type quasipotential equations. In Refs. \cite{Yin:2021uom,Gutierrez-Guerrero:2021rsx,Yu:2006ty,
Maris:2002yu}, the masses of the light diquarks are calculated by applying the Bethe-Salpeter equation. The potential model is used to obtain the mass spectra of diquarks in Ref. \cite{Ferretti:2019zyh}. In Ref. \cite{Hess:1998sd}, the diquark masses are obtained from lattice QCD.

The {\rt} is one of the effective approaches widely used in the study of hadron spectra \cite{Chew:1961ev,Regge:1959mz,Chew:1962eu,Nambu:1978bd,Ademollo:1969nx,Baker:2002km,
Afonin:2007jd,Dosch:2015nwa,Forkel:2007cm,Filipponi:1997vf,brau:04bs,Brisudova:1999ut,
Guo:2008he,Feng:2023ynf,Lovelace:1969se,Irving:1977ea,Collins:1971ff,
Inopin:1999nf,MartinContreras:2020cyg,Afonin:2023lfi,
MartinContreras:2022ebv,Patel:2022hhl,Abreu:2020ttf,Badalian:2019lyz,
FolcoCapossoli:2019imm,Chen:2014nyo,Afonin:2014nya,Chen:2023web,Chen:2018hnx,Chen:2018bbr,
Chen:2018nnr,Chen:2021kfw,Chen:2022flh,Chen:2023djq}. Although diquarks are colored states and not physical \cite{Jaffe:2004ph}, we previously attempted to apply the {\rt} approach to discuss the doubly heavy diquarks composed of two heavy quarks in Ref. \cite{Feng:2023txx}, and the heavy-light diquarks constituting of one heavy quark and one light quark in Ref. \cite{Chen:2023cws} by following the hadron {\rts} \cite{note}. The obtained results agree with the theoretical predictions calculated by using other methods.
In this study, we attempt to apply the {\rt} approach to investigate the light diquarks $(qq')$ composed of two light quarks. However, we find that the direct use of the {\rt} formula for light mesons and baryons fails, and therefore, the {\rt} formula needs to be modified to fit the {\rts} for the light diquarks $(ud)$, $(us)$, $(ss)$.

The paper is organized as follows: In Sec. \ref{sec:rgr}, the {\rt} relations are obtained from the spinless Salpeter equation (SSE). In Sec. \ref{sec:rtdiquark}, we investigate the {\rts} for the light diquarks. The conclusions are presented in Sec. \ref{sec:conc}.

\section{{\rt} relations for the light diquarks}\label{sec:rgr}

In this section, we attempt to seek unified {\rt} formula for the light mesons $q\bar{q}'$ ($q,q'=u,d,s$) and the light diquarks $(qq')$ [where $\bar{q}'$ is the antiquark of $q'$].

\subsection{SSE}
The {\sse} (SSE) \cite{Godfrey:1985xj,Ferretti:2019zyh,Durand:1981my,Durand:1983bg,Lichtenberg:1982jp,Jacobs:1986gv} reads as
\begin{eqnarray}\label{sse}
M\Psi_{d,m}({\bfr})=\left(\omega_1+\omega_2\right)\Psi_{d,m}({\bfr})+V_{d,m}\Psi_{d,m}({\bfr}),
\end{eqnarray}
where $M$ is the bound state mass (diquark or meson). $\Psi_{d,m}({\bfr})$ are the diquark wave function and the meson wave function, respectively. $V_{d,m}$ are the diquark potential and the meson potential, respectively, see Eq. (\ref{potential}). $\omega_1$ is the relativistic energy of quark $q$, and $\omega_2$ is of quark $q'$ or antiquark $\bar{q}'$,
\bea\label{omega}
\omega_i=\sqrt{m_i^2+{\bf p}^2}=\sqrt{m_i^2-\Delta}\;\; (i=1,2).
\eea
$m_1$ and $m_2$ are the effective masses of light quarks $q$ and $q'$ (or antiquark $\bar{q}'$), respectively.

According to the $SU_c(3)$ color symmetry, a meson is a color singlet composed of one quark in $3_c$ and one antiquark in {\cltb}. The diquark composed of two quarks in $3_c$ is a color antitriplet {\cltb} or a color sextet $6_c$.
Only the {\cltb} representation of $SU_c(3)$ is considered in the present work and the {\cls} representation \cite{Weng:2021hje,Praszalowicz:2022sqx} is not considered. %%
In $SU_c(3)$, there is attraction between quark pairs $(qq')$ in the color antitriplet channel and this is just twice weaker than in the color singlet $q\bar{q}'$ in the one-gluon exchange approximation \cite{Esposito:2016noz}.
It is introducing a factor $1/2$. One would expect $1/2$ to be a global factor since it comes from the color structure of the wavefunction and it is common to extend this factor to the whole potential describing the quark-quark interaction \cite{Debastiani:2017msn}.
Following Ref. \cite{Ferretti:2019zyh}, we employ the potential
\bea\label{potential}
V_{d,m}=-\frac{3}{4}\left[-\frac{4}{3}\frac{\alpha_s}{r}+{\sigma}r+C\right]
\left({\bf{F}_i}\cdot{\bf{F}_j}\right)_{d,m},
\eea
where $\alpha_s$ is the strong coupling constant of the color Coulomb potential. $\sigma$ is the string tension. $C$ is a fundamental parameter \cite{Gromes:1981cb,Lucha:1991vn}. The part in the bracket is the Cornell potential \cite{Eichten:1974af}. ${\bf{F}_i}\cdot{\bf{F}_j}$ is the color-Casimir,
\bea\label{mrcc}
\langle{(\bf{F}_i}\cdot{\bf{F}_j})_{d}\rangle=-\frac{2}{3},\quad
\langle{(\bf{F}_i}\cdot{\bf{F}_j})_{m}\rangle=-\frac{4}{3}.
\eea
The value of $({\bf{F}_i}\cdot{\bf{F}_j})_{d}$ is half of $({\bf{F}_i}\cdot{\bf{F}_j})_{m}$, which agrees with the relation
\cite{Faustov:2021hjs,Godfrey:1985xj,Debastiani:2017msn}
\bea\label{halfr}
V_{d}=\frac{V_{m}}{2}.
\eea
According to Eqs. (\ref{sse}), (\ref{potential}), and (\ref{halfr}), we see that the diquark and meson are described in an unified approach. Therefore, it is expected that the light diquarks and the light mesons should be described universally by the {\rt} approach.

\subsection{{\rts} relations}

The masses of the light quarks are assumed to approach zero, $m_1,m_2\to0$ in Refs. \cite{Chew:1962eu,Nambu:1978bd,Dosch:2015nwa} or is taken as being very small and then is considered by correction term in Refs. \cite{Selem:2006nd,Chen:2014nyo,Afonin:2014nya,Sonnenschein:2018fph}. In the limit $m_1,m_2\to0$, Eq. (\ref{sse}) is reduced to be
\begin{eqnarray}\label{sser}
M\Psi_{d,m}({\bfr})=2{\bfpa}\Psi_{d,m}({\bfr})+V_{d,m}\Psi_{d,m}({\bfr}).
\end{eqnarray}
we can obtain from Eq. (\ref{sser}) by employing the Bohr-Sommerfeld quantization approach \cite{Brau:2000st,brsom}
\begin{align}\label{rgll}
M{\sim}&2\sqrt{2\sigma'}\sqrt{l}\quad (l{\gg}n_r),\nonumber\\
M{\sim}&2\sqrt{{\pi}\sigma'}\sqrt{n_r}\quad (n_r{\gg}l),
\end{align}
where
\begin{eqnarray}\label{sigm}
\sigma'=\left\{\begin{array}{cc}
\frac{\sigma}{2}, & \text{diquark}, \\
\sigma, & \text{meson}.
\end{array}\right.
\end{eqnarray}
$l$ is the orbital angular momentum and $n_r$ is the radial quantum number.
From Eq. (\ref{rgll}) we can have the parameterized form \cite{Chen:2022flh}
\bea\label{massf}
M=m_R+\beta_x\sqrt{x+c_{0x}}\;\;(x=l,\,n_r)
\eea
with the parameters
\bea\label{massform}
\beta_x=c_{fx}c_xc_c,\;
c_c=\sqrt{\sigma'},\; c_l=2\sqrt{2},\; c_{n_r}=2\sqrt{\pi}.
\eea
All parameters are determined by fitting the meson {\rts} and can be used to calculate both the meson masses and the diquark masses. Both $c_{fl}$ and $c_{fn_r}$ are theoretically equal to 1 and are fitted in practice.
%%%
In Eq. (\ref{massform}), $c_x$ and $\sigma$ are universal for both the light mesons and the light diquarks. $c_{fx}$ and $c_{0x}$ are determined by fitting the given meson {\rt} and varies with different {\rts}.

For the light mesons and the light baryons, the common choice is \cite{Chew:1961ev,Afonin:2007jd,Chen:2022flh,Ishida:1994pf,Menapara:2022ksj},
\bea\label{mrfaa}
m_R=0.
\eea
This choice of $m_R$ leads to $M_d=\sqrt{1/2}M_m$, which is obtained from Eqs. (\ref{sigm}), (\ref{massf}), (\ref{massform}), and (\ref{mrfaa}). The predictions given by $M_d=\sqrt{1/2}M_m$ disagree with other theoretical results.
In reality, the light quarks are massive which should be included. According to Eqs. (\ref{potential}) and (\ref{sser}), $C$ should be also considered.
There are different ways to include the light quark mass correction \cite{Afonin:2014nya,Selem:2006nd,Chen:2014nyo,Sonnenschein:2018fph,
MartinContreras:2020cyg}.
In Ref. \cite{Afonin:2014nya}, the authors propose $M=m_1+m_2+\sqrt{a(n_r+{\alpha}l+b)}$ based on the string model, which differs from Eq. (\ref{massf}) with (\ref{mrfaa}) mainly in the term $m_1+m_2$.
According to the discussions in Refs. \cite{Afonin:2014nya,Chen:2022flh}, by including the parameter $C$, we try the modified formula Eq. (\ref{massf}) with
\bea\label{rtft}
m_R=m_1+m_2+C',
\eea
where
\begin{eqnarray}\label{mrcc}
C'=\left\{\begin{array}{cc}
\frac{C}{2}, & \text{diquark}, \\
C, & \text{meson}.
\end{array}\right.
\end{eqnarray}
As $m_1,m_2=0$ and $C$ is neglected, the modified formula  (\ref{massf}) with (\ref{rtft}) reduces to the usual {\rt} formula for the light mesons, i.e., (\ref{massf}) with (\ref{mrfaa}). (\ref{massf}) with (\ref{rtft}) has the same form as the {\rt} formula in Ref. \cite{Chen:2023cws}. And we note that while the former is applicable to the light systems, the latter is for the heavy-light systems.

The {\rt} relation for the doubly heavy diquarks has the same form as the {\rt} relation for the doubly heavy mesons \cite{Feng:2023txx}. However, unlike the case of doubly heavy diquarks and similar to the case of heavy-light diquarks \cite{Chen:2023cws}, the usual {\rt} formula (\ref{massf}) with (\ref{mrfaa}) for the light mesons cannot be directly applied to the light diquarks.
[Because $m_1+m_2+C'$ is small for $\beta_x\sqrt{x+c_{0x}}$ due to negative $C'$ and small $m_{1,2}$ and then $\beta_x\sqrt{x+c_{0x}}$ is a good approximation of $m_1+m_2+C'+\beta_x\sqrt{x+c_{0x}}$, the usual
Regge trajectory formula (\ref{massf}) with (\ref{mrfaa}) can work well for the light mesons.] The unified description of the light mesons and the light diquarks demands more rigorous Regge trajectory relations.
By fitting the masses of both light mesons and light diquarks, we find that the light diquark masses fitted by employing Eq. (\ref{massf}) with (\ref{massform}) and (\ref{mrfaa}) disagree with other theoretical predictions. However, by using the modified formula (\ref{massf}) with (\ref{massform}) and (\ref{rtft}), we can crudely estimate the masses of both light mesons and light diquarks, and these estimates are in agreement with other theoretical predictions, see details in the following section. Formulas (\ref{massf}) with (\ref{rtft}) includes not only the light quark masses $m_{1,2}$ but also the parameter $C$. Additionally, the half relation (\ref{halfr}) is explicitly included in formula by $\sigma'=\sigma/2$ and $C'=C/2$. The modified Regge trajectory formula provides an unified description of the light mesons and the light diquarks.

\begin{table*}[!phtb]
\caption{The completely antisymmetric states for the diquarks in {\cltb} and in $6_c$ \cite{Feng:2023txx}. $j$ is the spin of the diquark {\dqs}, $s$ denotes the total spin of two quarks, $l$ represents the orbital angular momentum. $n=n_r+1$, $n_r$ is the radial quantum number, $n_r=0,1,2,\cdots$. }  \label{tab:dqstates}
\centering
\begin{tabular*}{0.8\textwidth}{@{\extracolsep{\fill}}ccccc@{}}
\hline\hline
 Spin of diquark & Parity  &  Wave state  &  Configuration    \\
( $j$ )          & $(j^P)$ & $(n^{2s+1}l_j)$  \\
\hline
j=0              & $0^+$   & $n^1s_0$         & $[qq']^{{\cltba}}_{n^1s_0}$,\; $\{qq'\}^{{6_c}}_{n^1s_0}$ \\
                 & $0^-$   & $n^3p_0$         & $[qq']^{{\cltba}}_{n^3p_0}$,\; $\{qq'\}^{{6_c}}_{n^3p_0}$       \\
j=1              & $1^+$   & $n^3s_1$, $n^3d_1$   & $\{qq'\}^{{\cltba}}_{n^3s_1}$,\;    $\{qq'\}^{{\cltba}}_{n^3d_1}$,\;
$[qq']^{6_c}_{n^3s_1}$,\;    $[qq']^{{6_c}}_{n^3d_1}$\\
                 & $1^-$   & $n^1p_1$, $n^3p_1$   &
$\{qq'\}^{{\cltba}}_{n^1p_1}$,\; $[qq']^{{\cltba}}_{n^3p_1}$, \;          $[qq']^{6_c}_{n^1p_1}$,\; $\{qq'\}^{6_c}_{n^3p_1}$ \\
j=2              & $2^+$   & $n^1d_2$, $n^3d_2$         &  $[qq']^{{\cltba}}_{n^1d_2}$,\; $\{qq'\}^{{\cltba}}_{n^3d_2}$,\;
$\{qq'\}^{6_c}_{n^1d_2}$,\; $[qq']^{6_c}_{n^3d_2}$         \\
                 & $2^-$   & $n^3p_2$, $n^3f_2$       &
 $[qq']^{{\cltba}}_{n^3p_2}$,\; $[qq']^{{\cltba}}_{n^3f_2}$,\;
  $\{qq'\}^{6_c}_{n^3p_2}$,\; $\{qq'\}^{6_c}_{n^3f_2}$          \\
$\cdots$         & $\cdots$ & $\cdots$               & $\cdots$  \\
\hline\hline
\end{tabular*}
\end{table*}

\section{{\rts} for the light diquarks}\label{sec:rtdiquark}

In this section, the {\rts} for the light diquarks $(ud)$, $(us)$ and $(ss)$ are investigated.

\subsection{Preliminary}

When a diquark $(qq')$ contains light quarks with $q,q'=u,d,s$, the overall state function must be antisymmetric because strong interactions do not distinguish the flavor $u$, $d$, $s$ \cite{Esposito:2016noz}. The completely antisymmetric states for the diquarks in {\cltb} are listed in Table \ref{tab:dqstates}.
The state of diquark $(qq')$ is denoted as $[qq']^{\cltba}_{n^{2s+1}l_j}$ or $\{qq'\}^{\cltba}_{n^{2s+1}l_j}$. $\{qq'\}$ and $[qq']$ indicate the permutation symmetric and antisymmetric flavor wave functions, respectively. $n=n_r+1$, $n_r=0,1,\cdots$. $s$ is the total spin of two quarks, and $j$ is the spin of the diquark $(qq')$. ${\cltba}$ denotes the color antitriplet state of diquark.

The modified {\rt} formula (\ref{massf}) with (\ref{rtft}) is used to fit the radial and orbital {\rts} for the light diquarks.
%%%
The quality of a fit is measured by the quantity $\chi^2$ defined by \cite{Sonnenschein:2014jwa}
\bea
\chi^2=\frac{1}{N-1}\sum^{N}_{i=1}\left(\frac{M_{fi}-M_{ei}}{M_{ei}}\right)^2,
\eea
where $N$ is the number of points on the trajectory, $M_{fi}$ is fitted value and $M_{ei}$ is the experimental value or the theoretical value of the $i$-th particle mass. The parameters are determined by minimizing $\chi^2$.
%%%
Firstly, using the {\rt} formula (\ref{massf}) with (\ref{rtft}), the experimental data \cite{ParticleDataGroup:2022pth} and the theoretical data \cite{Godfrey:1985xj,Ebert:2009ub}, we fit the radial and orbital {\rts} for the light mesons, respectively. Secondly, we choose the following parameter values which are used to fit the {\rts} for the light diquarks,
\begin{align}\label{param}
m_{u,d}&=0.50\; {\gev},\quad m_s=0.67\; {\gev},\nonumber\\
\sigma&=0.18\; {\gev^2},\quad C=-0.64\; {\gev}.
\end{align}
These parameters are universal for all light diquark {\rts}. Thirdly, the parameters $c_{fx}$ and $c_{0x}$ in (\ref{massf}) are determined by fitting the corresponding meson {\rts}. For example, the $c_{fn_r}$ and $c_{0n_r}$ in the $\{ud\}^{\cltba}_{1^3s_1}$ Regge trajectory are calculated by fitting the radial Regge trajectory for the $\rho$ mesons. As all parameters are determined, the diquark masses can be calculated finally.
%%%
There is not compelling reason why $c_{0x}$ obtained by fitting the meson {\rts} can be directly used to calculate the diquark {\rts}. We use this method as a provisional method to determine $c_{0x}$ before finding a better one. It validates this method that the fitted results for the light diquarks $(ud)$, $(us)$ and $(ss)$ agree with the theoretical values obtained by using other approaches, see the discussions in the following subsections.

\begin{table}[!phtb]
\caption{The fitted values of parameters $c_{fn_r}$, $c_{fl}$, $c_{0n_r}$, and $c_{0l}$. $c_{fn_r}$ ($c_{fl}$) and $c_{0n_r}$ ($c_{0l}$) are obtained by fitting the radial (orbital) {\rts}. $1^1s_0$, $1^3s_1$ and $1^3p_0$ mean that the parameters are calculated by fitting the {\rts} for the $1^1s_0$ state, the $1^3s_1$ state and the $1^3p_0$ state, respectively.}  \label{tab:fitparameters}
\centering
\begin{tabular*}{0.47\textwidth}{@{\extracolsep{\fill}}cccc@{}}
\hline\hline
                   & $(ud)$ &  $(us)$  & $(ss)$  \\
\hline
$c_{fn_r}$ ($1^1s_0$)  &0.668  &0.683  &    \\
$c_{0n_r}$ ($1^1s_0$)  &0.0    &0.005  &    \\
$c_{fn_r}$ ($1^3s_1$)  &0.691  &0.708  &0.676    \\
$c_{0n_r}$ ($1^3s_1$)  &0.16   &0.12   &0.095    \\
$c_{fl}$  ($1^1s_0$)  &0.731   &0.714  &0.737    \\
$c_{0l}$ ($1^1s_0$)   &0.02    &0.005  &0.0    \\
$c_{fl}$  ($1^3s_1$)  &0.775   &0.734  &0.761    \\
$c_{0l}$ ($1^3s_1$)   &0.19    &0.165  &0.11    \\
$c_{fl}$  ($1^3p_0$)  &0.693   &0.695  &    \\
$c_{0l}$ ($1^3p_0$)   &0.0     &0.0    &    \\
\hline
\hline
\end{tabular*}
\end{table}

\subsection{{\rts} for the $(ud)$ diquark}

\begin{table}[!phtb]
\caption{The fitted values (in {\gev}) for the diquarks $(ud)$, $(us)$, and $(ss)$ by using the radial {\rts}. $n=n_r+1$, $n_r=0,1,\cdots$. $n_r$ is the radial quantum number. $s$ is the total spin of two quarks, $l$ is the orbital quantum number and $j$ is the spin of diquark. $\times$ denotes the nonexistent states.}  \label{tab:rad}
\centering
\begin{tabular*}{0.47\textwidth}{@{\extracolsep{\fill}}cccc@{}}
\hline\hline
State ($n^{2s+1}l_j$) & $(ud)$ &  $(us)$  & $(ss)$   \\
\hline
$1^1s_0$           & 0.68    & 0.90   &  $\times$ \\
$2^1s_0$           & 1.39    & 1.58   &  $\times$ \\
$3^1s_0$           & 1.68    & 1.88   &  $\times$ \\
$4^1s_0$           & 1.91    & 2.11   &  $\times$ \\
$5^1s_0$           & 2.10    & 2.30   &  $\times$ \\
%%%
$1^3s_1$           & 0.97    & 1.11   & 1.24    \\
$2^3s_1$           & 1.47    & 1.65   & 1.77    \\
$3^3s_1$           & 1.76    & 1.95   & 2.06   \\
$4^3s_1$           & 1.99    & 2.18   & 2.28   \\
$5^3s_1$           & 2.18    & 2.38   & 2.47   \\
\hline
\hline
\end{tabular*}
\end{table}

\begin{table}[!phtb]
\caption{Same as Table \ref{tab:rad} except by using the orbital {\rts}. }  \label{tab:orb}
\centering
\begin{tabular*}{0.47\textwidth}{@{\extracolsep{\fill}}cccc@{}}
\hline\hline
State ($n^{2s+1}l_j$) & $(ud)$ &  $(us)$  & $(ss)$  \\
\hline
$1^1s_0$         & 0.77    & 0.89   & 1.02($\times$) \\
$1^1p_1$         & 1.31    & 1.46   & 1.65 \\
$1^1d_2$         & 1.56    & 1.71   & 1.90 ($\times$) \\
$1^1f_3$         & 1.76    & 1.90   & 2.10 \\
$1^1g_4$         & 1.92    & 2.06   & 2.27 ($\times$)\\
$1^1h_5$         & 2.07    & 2.21   & 2.42 \\
%%%\hline
$1^3s_1$         & 0.97    & 1.10   & 1.23 \\
$1^3p_2$         & 1.40    & 1.52   & 1.70 ($\times$)\\
$1^3d_3$         & 1.65    & 1.77   & 1.96 \\
$1^3f_4$         & 1.85    & 1.96   & 2.16 ($\times$)\\
$1^3g_5$         & 2.03    & 2.12   & 2.33 \\
$1^3h_6$         & 2.18    & 2.27   & 2.48 ($\times$)\\
\hline\hline
\end{tabular*}
\end{table}

\begin{table*}[!phtb]
\caption{Comparison of theoretical predictions for the masses of the light diquarks (in {\gev}).}  \label{tab:cpmass}
\centering
\begin{tabular*}{0.8\textwidth}{@{\extracolsep{\fill}}ccccccc@{}}
\hline\hline
 $j^P$ & Diquark  &Our results & FGS \cite{Faustov:2021hjs}, FG \cite{Faustov:2015eba}   & F \cite{Ferretti:2019zyh}
 & YCRS \cite{Yin:2021uom}  & GTB \cite{Gutierrez-Guerrero:2021rsx}
  \\
\hline
$0^+$   &  $[ud]^{{\cltba}}_{1^1s_0}$
              & 0.68   &0.710  &0.691    &0.78     &0.77      \\
        &  $[ud]^{{\cltba}}_{2^1s_0}$
              & 1.39   &1.513  &        &          &      \\
        &  $[us]^{{\cltba}}_{1^1s_0}$
              & 0.90   &0.948  &0.886    & 0.93    &0.92       \\
$0^-$   & $[ud]^{{\cltba}}_{1^3p_0}$
             & 1.27   &1.321  &          &1.15     &1.30       \\
        & $[us]^{{\cltba}}_{1^3p_0}$
             & 1.44   &       &          &1.28     &1.41      \\
%%%
$1^+$   & $\{ud\}^{{\cltba}}_{1^3s_1}$
             & 0.97   &0.909  & 0.840   &1.06     &1.06       \\
        & $\{ud\}^{{\cltba}}_{2^3s_1}$
             & 1.47   &1.630  &         &         &         \\
        & $\{us\}^{{\cltba}}_{1^3s_1}$
             & 1.11   &1.069  &0.992    &1.16     &1.16       \\
        & $\{ss\}^{{\cltba}}_{1^3s_1}$
             &1.24    &1.203  &1.136    &1.26     &1.25       \\
        & $\{ss\}^{{\cltba}}_{2^3s_1}$
             &1.77    &1.817  &         &     &      \\
$1^-$   & $\{ud\}^{{\cltba}}_{1^1p_1}$
             &1.31    &1.397  &         &1.33     &1.44       \\
        & $\{us\}^{{\cltba}}_{1^1p_1}$
             &1.46    &       &         &1.44     &1.54       \\
        & $\{ss\}^{{\cltba}}_{1^1p_1}$
            & 1.65   &1.608  &          &         &1.64     \\
\hline\hline
\end{tabular*}
\end{table*}

\begin{figure*}[!phtb]
\centering
\subfigure[]{\label{figudr}\includegraphics[scale=0.75]{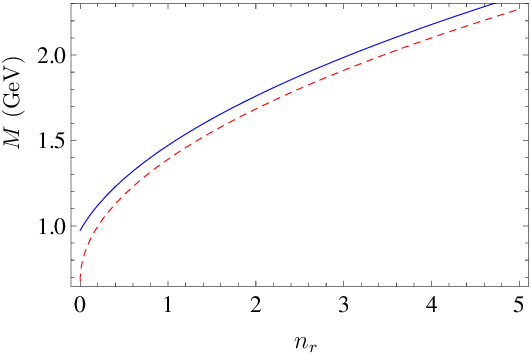}}
\subfigure[]{\label{figudo}\includegraphics[scale=0.75]{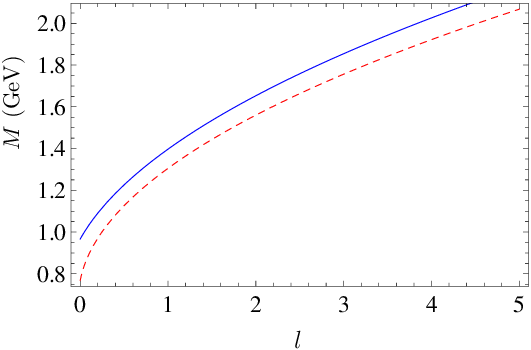}}
\caption{The radial and orbital {\rts} for the $(ud)$ diquark. (a) Radial {\rts} for the $[ud]^{\bar{3}_c}_{1^1s_0}$ state (the red dashed line) and the $\{ud\}^{\bar{3}_c}_{1^3s_1}$ state (the blue line). (b) Orbital {\rts} for the $[ud]^{\bar{3}_c}_{1^1s_0}$ state (the red dashed line) and the $\{ud\}^{\bar{3}_c}_{1^3s_1}$ state (the blue line). The data are listed in Tables \ref{tab:fitparameters}, \ref{tab:rad} and \ref{tab:orb}.}\label{fig:ud}
\end{figure*}

Using Eq. (\ref{massf}) with (\ref{rtft}) to fit the radial {\rts} for the $\pi$ mesons and for the $\rho$ mesons, respectively, we obtain the parameters $c_{fn_r}$ and $c_{0n_r}$, see Table \ref{tab:fitparameters}. The experimental data from PDG \cite{ParticleDataGroup:2022pth} and the theoretical data from \cite{Ebert:2009ub} are used to obtain $c_{fn_r}$ and $c_{0n_r}$. As we fit the radial ${\pi}$ {\rt}, the first point, i.e., $1^1S_0$ state is excluded due to its abnormally small mass. Substitute the values in Eq. (\ref{param}) and the obtained $c_{fn_r}$ and $c_{0n_r}$ into (\ref{massf}), (\ref{massform}), (\ref{rtft}), and (\ref{mrcc}). Then the masses of diquark $(ud)$ can be calculated by Eq. (\ref{massf}) with (\ref{rtft}), see Table \ref{tab:rad}. The radial {\rts} are shown in Fig. \ref{figudr}.

Similar to the radial {\rt} case, the orbital $\pi(1^1S_0)$ and $\rho(1^3S_1)$ {\rts} are fitted by using Eq. (\ref{massf}) with (\ref{rtft}), respectively. Using the experimental data from PDG \cite{ParticleDataGroup:2022pth} and the theoretical data from \cite{Ebert:2009ub}, $c_{fn_r}$ and $c_{0n_r}$ can be fitted. As the orbital ${\pi}$ {\rt} is fitted, the first point is also excluded. Substitute the values in Eq. (\ref{param}) and the fitted $c_{fn_r}$ and $c_{0n_r}$ into (\ref{massf}), (\ref{massform}), (\ref{rtft}), and (\ref{mrcc}). Then the masses of diquark $(ud)$ can be calculated by Eq. (\ref{massf}) with (\ref{rtft}), see Table \ref{tab:orb}. The orbital {\rts} are shown in Fig. \ref{figudo}.

The calculated light diquark masses by using the {\rts} are in accordance with other theoretical predictions, see Table \ref{tab:cpmass}. In Table \ref{tab:cpmass}, the mass of $[ud]^{{\cltba}}_{1^3p_0}$ is calculated by fitting the the orbital $a_0(1^3P_0)$ {\rt}. The mass of $\{ud\}^{{\cltba}}_{1^1p_1}$ is obtained by fitting the orbital ${\pi}$ {\rt} and is taken from Table \ref{tab:orb}.

\subsection{{\rts} for the $(us)$ diquark}

\begin{figure*}[!phtb]
\centering
\subfigure[]{\label{figusr}\includegraphics[scale=0.75]{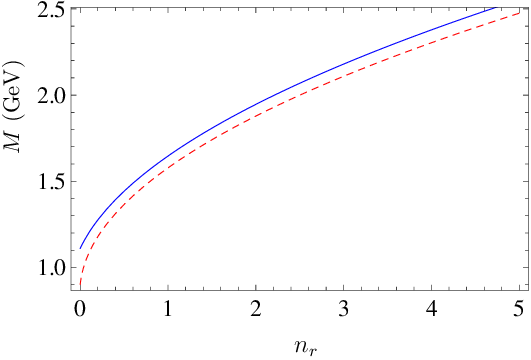}}
\subfigure[]{\label{figuso}\includegraphics[scale=0.75]{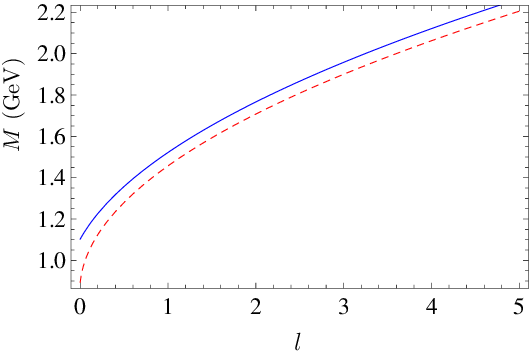}}
\caption{Same as Fig. \ref{fig:ud} except for the $(us)$ diquark.}\label{fig:us}
\end{figure*}

Using Eq. (\ref{massf}) with (\ref{rtft}) to fit the radial {\rts} for the $K$ mesons and for the $K^{\ast}$ mesons, respectively, we obtain the parameters $c_{fn_r}$ and $c_{0n_r}$, see Table \ref{tab:fitparameters}. The experimental data from PDG \cite{ParticleDataGroup:2022pth} and the theoretical data from \cite{Godfrey:1985xj,Ebert:2009ub} are used to obtain $c_{fn_r}$ and $c_{0n_r}$. Substitute the values in Eq. (\ref{param}) and the obtained $c_{fn_r}$ and $c_{0n_r}$ into (\ref{massf}), (\ref{massform}), (\ref{rtft}), and (\ref{mrcc}). Then the masses of diquark $(us)$ can be calculated by Eqs. (\ref{massf}) with (\ref{rtft}), see Table \ref{tab:rad}. The radial {\rts} are shown in Fig. \ref{figusr}.

Similar to the radial {\rt} case, the orbital $K(1^1S_0)$ and $K^{\ast}(1^3S_1)$ {\rts} are fitted by using Eq. (\ref{massf}) with (\ref{rtft}). Using the experimental data from PDG \cite{ParticleDataGroup:2022pth} and the theoretical data from \cite{Ebert:2009ub}, $c_{fn_r}$ and $c_{0n_r}$ can be fitted. Substitute the values in Eq. (\ref{param}) and the fitted $c_{fn_r}$ and $c_{0n_r}$ into (\ref{massf}), (\ref{massform}), (\ref{rtft}), and (\ref{mrcc}). Then the masses of diquark $(us)$ can be calculated by Eq. (\ref{massf}) with (\ref{rtft}), see Table \ref{tab:orb}. The orbital {\rts} are shown in Fig. \ref{figuso}.

The calculated $(us)$ masses by using the {\rts} are in accordance with other theoretical predictions, see Table \ref{tab:cpmass}. In Table \ref{tab:cpmass}, the mass of $[us]^{{\cltba}}_{1^3p_0}$ is calculated by fitting the the orbital $K_0(1^3P_0)$ {\rt}. The mass of $\{us\}^{{\cltba}}_{1^1p_1}$ is obtained by fitting the orbital ${K}$ {\rt} and is taken from Table \ref{tab:orb}.

In this study, we do not consider the $^1l_l-^3l_l$ mixing of the strange diquarks $(us)$ which is similar to the mixing of spin-triplet and spin-singlet states of the strange mesons \cite{Ebert:2009ub}. The masses of the $1^1l_l$ states of the strange diquarks are estimated by using the orbital {\rts} for the $[us]^{\cltba}_{1^1s_0}$ state, see Table \ref{tab:orb}.

\subsection{{\rts} for the $(ss)$ diquark}

\begin{figure*}[!phtb]
\centering
\subfigure[]{\label{figssr}\includegraphics[scale=0.75]{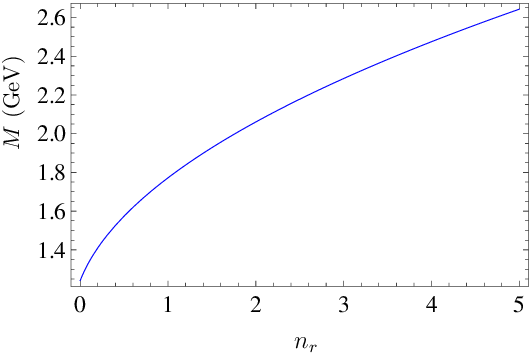}}
\subfigure[]{\label{figsso}\includegraphics[scale=0.75]{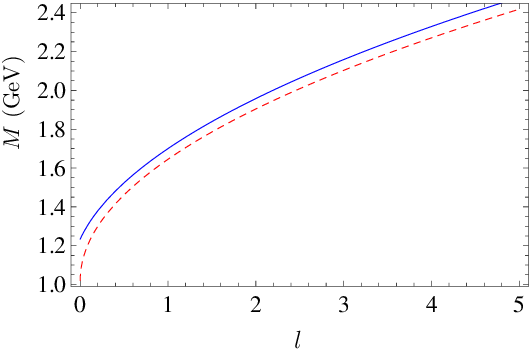}}
\caption{Same as Fig. \ref{fig:ud} except for the $(ss)$ diquark.}\label{fig:ss}
\end{figure*}

Because the diquark $(ss)$ composed of two identical quarks, the
antisymmetric flavor wave function does not exist; therefore, the states in the $[ss]$ configuration disappear in Table \ref{tab:dqstates}.
Using Eq. (\ref{massf}) with (\ref{rtft}) to fit the radial {\rts} for the $\varphi$ mesons, we obtain the parameters $c_{fn_r}$ and $c_{0n_r}$, see Table \ref{tab:fitparameters}. The experimental data from PDG \cite{ParticleDataGroup:2022pth} and the theoretical data from \cite{Ebert:2009ub} are used to obtain $c_{fn_r}$ and $c_{0n_r}$. Substitute the values in Eq. (\ref{param}) and the obtained $c_{fn_r}$ and $c_{0n_r}$ into (\ref{massf}), (\ref{massform}), (\ref{rtft}), and (\ref{mrcc}). Then the masses of diquark $(ss)$ can be calculated by Eq. (\ref{massf}) with (\ref{rtft}), see Table \ref{tab:rad}. The radial {\rt} for the $[ss]^{{\cltba}}_{1^1s_0}$ does not exist. The radial {\rt} for the $[ss]^{{\cltba}}_{1^3s_1}$ is shown in Fig. \ref{figssr}.

Similar to the radial {\rt} case, the orbital $\eta(1^1S_0)$ and $\varphi(1^3S_1)$ {\rts} are fitted by using Eq. (\ref{massf}) with (\ref{rtft}). Using the experimental data from PDG \cite{ParticleDataGroup:2022pth} and the theoretical data from \cite{Ebert:2009ub}, $c_{fn_r}$ and $c_{0n_r}$ can be fitted. Substitute the values in Eq. (\ref{param}) and the fitted $c_{fn_r}$ and $c_{0n_r}$ into (\ref{massf}), (\ref{massform}), (\ref{rtft}), and (\ref{mrcc}). Then the masses of diquark $(ss)$ can be calculated by Eqs. (\ref{massf}) and (\ref{rtft}), see Table \ref{tab:orb}. The orbital {\rts} are shown in Fig. \ref{figsso}. Due to the pauli exclusion principle, some states do not exist, see Table \ref{tab:orb}.

The calculated $(ss)$ masses by using the {\rts} are in accordance with other theoretical predictions, see Table \ref{tab:cpmass}. In Table \ref{tab:cpmass}, the mass of $\{ss\}^{{\cltba}}_{1^1p_1}$ is obtained by fitting the orbital $\eta(1^1S_0)$ {\rt} and is taken from Table \ref{tab:orb}.

\subsection{Discussions}

When using diquark in multiquark systems, the interactions between quark and quark, diquark and quark, diquark and diquark are needed. In Refs. \cite{Faustov:2021hjs,Faustov:2015eba,Ebert:2011kk,Ebert:2005nc}, these interactions are constructed with the help of the off-mass-shell scattering amplitude, which is projected onto the positive energy states.
The interactions also can be established by expanding the interactions of the quark-antiquark system to the quark-quark system, and then to the diquark-antidiquark systems or the diquark-quark systems \cite{Lundhammar:2020xvw}.
Furthermore, the effect of the finite size of diquark is treated differently. In Refs. \cite{Faustov:2021hjs,Faustov:2015eba,Ebert:2011kk,Ebert:2005nc}, the size of diquark is taken into account through corresponding form factors. At times, diquark is taken as being pointlike \cite{Ferretti:2019zyh,Lundhammar:2020xvw}, and we use this approximation in the present work.

In Ref. \cite{Chen:2023cws} and this work, it is shown that in order to describe the meson $q\bar{q}'$ and the corresponding diquark $(qq')$ universally, the masses of the light quarks should be taken into account in the Regge trajectories for both the light diquarks and the heavy-light diquarks.
Additionally, we find that the inclusion of a negative parameter is necessary in the diquark Regge trajectory formula according to the discussions in Refs. \cite{Feng:2023txx,Chen:2023cws} and this work.
Specifically, the fundamental parameter $C$ in the Cornell potential $-(4/3)(\alpha_s/r)+{\sigma}r+C$ is emphasized and becomes essential when discussing the {\rts} for various types of diquarks (including doubly heavy diquarks, heavy-light diquarks, and light diquarks). Without considering the parameter $C$, the meson $q\bar{q}'$ and the corresponding diquark $(qq')$ cannot be universally described using the {\rt} approach.

In Ref. \cite{Feng:2023txx}, we show an universal description of the {\rts} for the doubly heavy mesons and the doubly heavy diquarks. In Ref. \cite{Chen:2023cws}, we present the universal description of the {\rts} for the heavy-light mesons and the heavy-light diquarks. In this work, we show the universal description of the {\rts} for the light mesons and the light diquarks.
However, there is a limitation to the unified description of the light mesons and the light diquarks. The values of $C$ and the quark masses in the {\rts} for the doubly heavy diquark and for the heavy-light diquark are universal \cite{Feng:2023txx,Chen:2023cws}. However, in the case of light diquarks, the parameters in equation (\ref{param}) are only universal for the {\rts} of light diquarks. Furthermore, the fitted parameters $c_{fn_r}$ and $c_{fl}$ vary for different {\rts} in the light diquark case, which is different from the cases of doubly heavy diquarks and heavy-light diquarks. Therefore, it is expected that the {\rt} formula (\ref{massf}) with (\ref{rtft}) serves as a provisional formula until a better one is found.

\section{Conclusions}\label{sec:conc}
We attempt to apply the Regge trajectory approach to investigate the light diquarks constituting of two light quarks. The spectra of the light diquarks $(ud)$, $(us)$ and $(ss)$ are crudely estimated by the {\rt} approach, and these results are in agree with other theoretical results. This demonstrates the suitability of the {\rt} method for studying light diquarks.
The diquark Regge trajectory offers a new and very simple approach for estimating the spectra of the light diquarks.

The {\rt} relation for doubly heavy diquarks follows the same form as the {\rt} relation for doubly heavy mesons. However, unlike the case of doubly heavy diquarks, the usual {\rt} formula for light mesons cannot be directly applied to light diquarks. Instead, similar to the case of heavy-light diquarks, we need to consider the light quark mass and the parameter $C$ in the Cornell potential to obtain agreeable results. By incorporating these factors, a modified {\rt} formula is proposed, which has the same form as the formula for the heavy-light diquarks. The modified {\rt} formula provides an unified description of both light mesons and light diquarks.

We present a method to determine the parameters in the diquark {\rts}. By employing  (\ref{massf}) with (\ref{rtft}) to fit the light mesons, we can obtain values of the universal parameters. By fitting a chosen meson {\rt}, $c_{fx}$ and $c_{0x}$ are calculated. Once all parameters are computed, the light diquark {\rt} is definite and the spectra of the light diquarks can be estimated.

Diquarks are not physical objects and their spectrum cannot be obtained experimentally. However, in the diquark picture, the excited states of diquarks can be identified by the $\rho-$mode excited states of baryons, tetraquarks and pentaquarks. For instance, the behavior of the $\rho-$mode excited states of the doubly heavy baryons with respect to $l$ or $n_r$ will be different from the behavior of the $\lambda-$mode excited states. This can be tested in future experiments.
It is expected that the diquark Regge trajectory will provide a simple method for investigating the $\rho-$mode excitations of baryons, tetraquarks and pentaquarks containing the light diquarks.

\section*{Acknowledgments}
We are very grateful to the anonymous referees for the
valuable comments and suggestions.


\begin{thebibliography}{99}
\bibitem{note}
The {\rts} are commenly plotted in the $(M^2,\,x)$ plane or in the $(x,\,M^2)$ plane, where $x=l,\,n_r$. For simplicity, the figures plotted in the $(M,\,x)$ plane, the $(M-m_R,\,x)$ plane and the $((M-m_R)^2,\,x)$ plane are also called the {\rts}.
\end{thebibliography}
\end{document}